\documentclass{pnastwo_modified}
\usepackage{amsmath,amsfonts,amssymb,graphicx}

\newcommand{\eps}{\varepsilon}
\newcommand{\beqref}[1]{{\bf\ref{#1}}}

\newcommand{\del}{\partial}

\DeclareMathOperator{\tr}{tr}
 
\providecommand{\norm}[1]{\lVert#1\rVert}

\begin{document}

\title{Network synchronization landscape reveals compensatory structures, quantization, and 
the~positive effect of negative interactions}

\author{%
Takashi Nishikawa%
\affil{1}{Department of Physics and Astronomy, Northwestern University, Evanston, IL 60208, USA}%
\affil{2}{Department of Mathematics, Clarkson University, Potsdam, NY 13699, USA}%
\and
Adilson E. Motter%
\affil{1}{}%
\affil{3}{Northwestern Institute on Complex Systems, Northwestern University, Evanston, IL 60208, 
USA}
}

\contributor{Published in PNAS 107, 10342-10347 (2010)}

\maketitle

\begin{article}
\begin{abstract} 
Synchronization, in which individual dynamical units keep in pace with each other in a decentralized 
fashion, depends both on the dynamical units and on the properties of the interaction network.
Yet, the role played by the network has resisted comprehensive characterization
within the prevailing paradigm that
interactions facilitating pair-wise synchronization also facilitate collective synchronization. 
Here we challenge this paradigm and show that networks with best complete synchronization, least 
coupling cost, and maximum dynamical robustness, have arbitrary complexity but {\em quantized} 
total interaction strength, which constrains the allowed number of connections. 
It stems from this characterization that {\em negative} interactions as well as link {\em removals} can 
be used to systematically improve and optimize synchronization properties in both directed and 
undirected networks.
These results extend the recently discovered compensatory perturbations in metabolic networks to the 
realm of oscillator networks and demonstrate why ``less can be more" in network synchronization.
\end{abstract}

\begin{figure}[t]
\begin{center}
\includegraphics[width=0.4\textwidth]{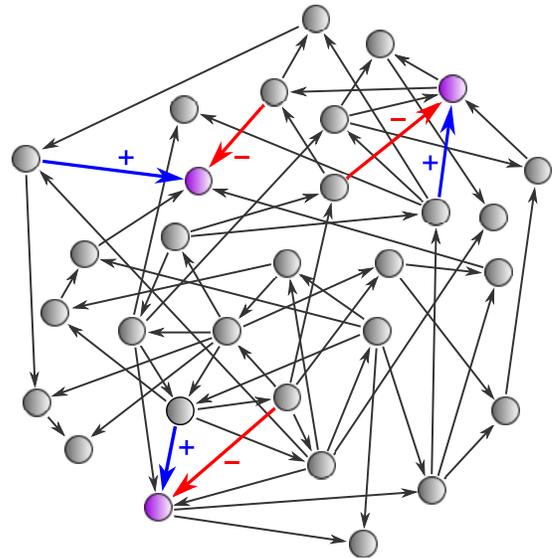}
\end{center}
\caption{\label{ExampleOptimal}
Examples of compensatory network structures.
Nodes represent identical dynamical units and links diffusive couplings.
The network consisting of the black and blue links, all having strength $+1$, is not an optimal network 
for synchronization, namely one that synchronizes over the widest range of global coupling strength (to be formalized in Results section).
However, it can be made optimal either by 1) removing the blue links, 2) adding the red links with 
negative strength $-1$, or 3) scaling the strengths of all in-links of the purple nodes by a factor of 2/3.
}
\end{figure}

\dropcap{F}locking animals \cite{T.-Vicsek:1995dw,Cucker:2007fc}, self-coordinated moving sensors 
\cite{P.-Yang:2008ay}, bursting neurons \cite{Izhikevich:2001ss}, pace-matching chemical oscillators 
\cite{Taylor:2009yq}, and frequency-locked power generators \cite{Grainger:1994yo} are some of 
many physical manifestations of spontaneous synchronization.
Like other forms of collective phenomena \cite{A.-Barrat:2008qc}, synchronization depends  critically 
on the properties of the interaction network \cite{Strogatz:2001il}.
Common wisdom suggests that synchronization is generally easier to achieve with more interactions, 
that synchronization properties change monotonically as the number of available interactions is varied, 
and that certain network structures facilitate while others inhibit synchronization.  
These three expectations, however, are all false because they ignore 
the possibility of compensatory structural effects. 
For example, removing a link from a globally connected network 
makes it less likely to synchronize,
but targeted removal of additional links can 
enhance synchronization \cite{Nishikawa:2006fk,Nishikawa:2006kx,Yin:2006hs,Duan:2007ve,Gu:2009kx,Hagberg:2008wd,1674-1056-18-8-004}.
Heterogeneous distribution of coupling strengths or connectivity generally inhibits synchronization 
\cite{Nishikawa:2003xr,Zhou:2006wb}, 
but  when combined they can compensate for each other \cite{Motter2004af,Motter:2005ub}.
Bringing this argument one step further, 
while previous studies have focused mainly on positive interactions (but see 
Refs.~\cite{Restrepo:2006fk,restrepo:015107,0295-5075-89-2-20002}) ---
presumably because negative interactions alone generally do not support synchrony --- it is actually 
easy to provide examples in which
negative interactions help stabilize synchronous states. 
This is illustrated in Fig.~\ref{ExampleOptimal}, where the network composed of black and blue links
is not 
optimal for synchronization
but the removal of the blue interactions or, alternatively, the addition of 
interactions with negative strengths (red links) makes it optimal;  
the same is achieved by weighting the strengths of all three input interactions of each purple node by a 
factor of $2/3$.
However, the counter-intuitive properties that start to emerge from such case studies currently lack a 
common in-depth explanation that is both comprehensive and predictive in nature.

Here we show that these and other apparently disparate properties follow from the discovery we 
present below that networks optimal
for synchronization
have a quantized number of links, in multiples of a constant that depends only on the number of nodes 
and on the connection strengths.
We derive our results for the local stability of synchronous states in networks of identical units and 
provide evidence that our main results remain valid for networks of non-identical units.
We choose to focus on optimal networks because, as we show, this class is very rich, can be dealt with 
analytically, 
and forms a multi-cusped synchronization landscape, which underlies all synchronizable networks and 
from which suboptimal networks can be studied using perturbation and numerical methods.
An immediate consequence of our quantization result is that the addition of links to an optimal network 
generally results in a suboptimal network, providing systematic examples of link removals that enhance 
synchronizability.
Similar quantization results hold true for optimal networks with negative interactions, which we derive using 
a generalized complement transformation that maps them into networks with only positive 
interactions. 
We show that negative interactions can reveal antagonistic structural relations and counterbalance structural heterogeneities, with potential implications for inhibitory interactions in neural 
\cite{Bragin:1995uq,Whittington-MA:1995yq,Ylinen:1995kx}, power-grid~\cite{Grainger:1994yo,Myers:2009fk}, and cell-regulatory networks \cite{McAdams:2003sy}.  The interactions 
between power-grid generators, for example, can be positive or negative depending on the inductive 
versus capacitive nature of the corresponding network elements 
(e.g., for the Western U.S.\ power grid 
they are split 97\% versus 3\%~\cite{Myers:2009fk}).

\section*{Results}
\subsection*{Optimal networks for synchronization.}
We represent the structure of a network with $n$ nodes using its adjacency matrix $A = (A_{ij})_{1 
\le i,j \le n}$, where $A_{ij}$ 
is the strength of the link from node $j$ to node $i$.
We consider the network dynamics governed by
\begin{align}\label{MainEqn}
\dot{\mathbf{x}}_i 
&= \mathbf{F}(\mathbf{x}_i) + \bar\varepsilon \sum_{j=1}^n A_{ij}[\mathbf{H}(\mathbf{x}_j) - 
\mathbf{H}(\mathbf{x}_i)], 
\end{align}
where $\mathbf{x}_i$ is the state vector of the $i$th dynamical unit, $\mathbf{F}$ represents the 
dynamics of an isolated unit, $\mathbf{H}(\mathbf{x}_j)$ is the signal that the $j$th unit sends to 
other units~\cite{Pecora:1998zp}, and $\bar\varepsilon = \varepsilon/d$ is the global coupling 
strength $\varepsilon \ge 0$ normalized by the the average coupling strength per node 
$d := \frac{1}{n}\sum_i\sum_{j \neq i} A_{ij}$.
As a result of this normalization,
the dynamics for a given $\varepsilon$ is
invariant 
under scaling of $A$ by a constant, which does not change the network structure.
This system has served as a workforce model to study synchronization because it allows analysis of the 
network influence without detailed specification of the properties of the dynamical units.
For example, using a 
stability function $\Lambda(\cdot)$ that is independent of the network structure \cite{A.-Barrat:2008qc,Pecora:1998zp,Dorogovtsev:2008ly,Arenas:2008yq},
the condition for a synchronous state 
${\bf x}_1(t) = \cdots = {\bf x}_n(t) = \mathbf{s}(t)$
to be linearly stable is $\Lambda(\bar\varepsilon\lambda_i) < 0$ for $i = 2,\ldots,n$, where $
\lambda_2, \ldots, \lambda_n$ are the non-identically zero 
eigenvalues of the Laplacian matrix $L$ defined by $L_{ij} = \delta_{ij} \sum_k A_{ik} - 
A_{ij}$ (see {\it Materials and Methods}).
Thus, 
the smaller 
the normalized spread of the eigenvalues in the complex plane, which we quantify using
\begin{equation}\label{eqn:sigma}
\sigma^2 := \frac{1}{d^2(n-1)} \sum_{i=2}^n |\lambda_i - \bar{\lambda}|^2, 
\text{ where }
\bar{\lambda} :=  \frac{1}{n-1} \sum_{i=2}^n \lambda_i,
\end{equation}
the more synchronizable the network will generally be.
Another measure of synchronizability is the coupling cost 
$K := \bar\eps_s \sum_i\sum_{j \neq i} A_{ij} = \bar\eps_s \cdot m$
at the synchronization threshold $\bar\eps = \bar\eps_s$,
whose minimization is equivalent to the condition
\begin{equation} \label{Optimal}
\lambda_2 = \lambda_3 = \cdots = \lambda_n = \bar\lambda > 0,
\end{equation}
which is also equivalent to the condition $\sigma = 0$.
This condition constrains all the eigenvalues to be real, even though the networks can be directed and 
directed networks have complex eigenvalues in general.
Condition~\beqref{Optimal} is
also equivalent to the maximization of the range of $\bar\eps$ that allows for stable synchronization as 
well as to the maximization of dynamical robustness, in that it can achieve the fastest exponential 
convergence to synchronization (see {\it Materials and Methods}).
(This may relate, for example, to the finding that heterogeneous distribution of links, which tend to 
make the distribution of $\lambda_i$'s heterogeneous, leads to longer transients to synchronization in 
ecological network models with strong couplings \cite{Holland:2008oq}.)
We emphasize that the equivalence of these conditions holds for a wide variety of possible stability 
functions, including those for which the region defined by $\Lambda(\cdot) < 0$ may be finite or semi-
infinite, or have multiple disconnected components.
Having the maximum range, minimum coupling cost, and maximum dynamical robustness, 
the networks satisfying Eq.~\beqref{Optimal} are called the optimal networks for 
synchronization.
Similar results can be derived for networks of non-identical units, in which the functions $\mathbf{F}$ and $\mathbf{H}$ are possibly different for different nodes, and this more general case will be discussed below.

\subsection*{Quantized number of links.}
\begin{figure}[t]
\begin{center}
\includegraphics[width=3in]{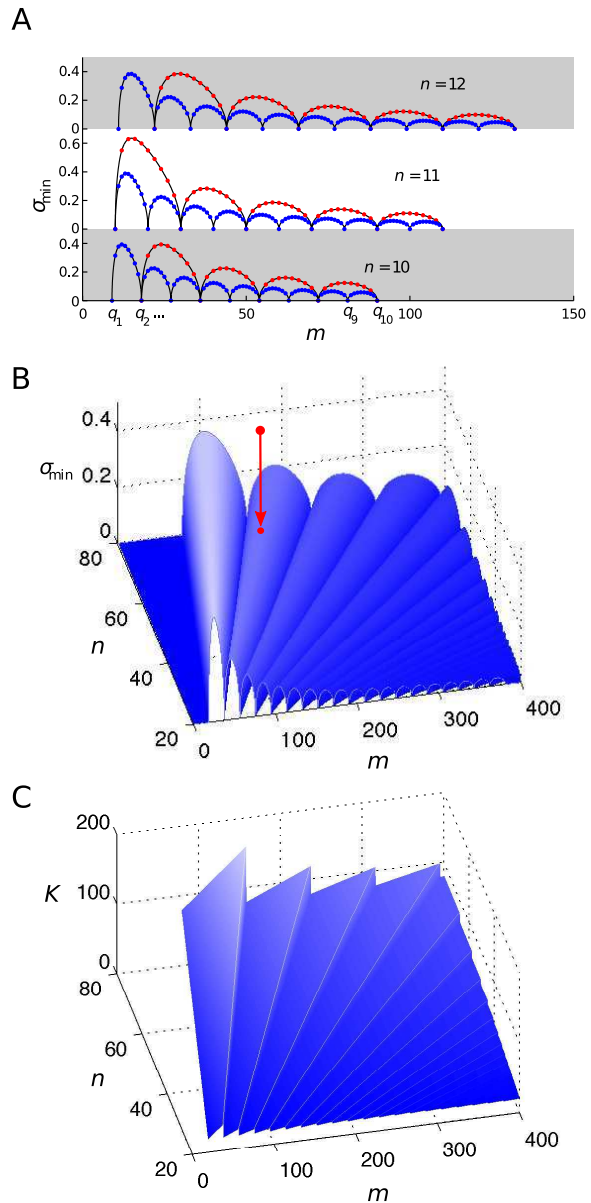}
\end{center}
\caption{\label{Quantization}
Quantization in the synchronization and cost landscapes.
(\textit{A}) Estimated by simulated annealing for networks with $n=10$--$12$ nodes, 
the minimum value $\sigma_{\min}$ (blue dots) of $\sigma$ (Eq.~\beqref{eqn:sigma})  shows cusp 
points at periodic values of the total interaction strength, $m = q_k$ (Eq.~\beqref{quantization}).
In this representation as a function of $m$,
the results for networks with positive ($A_{ij} = 0, 1$) and mixed positive/negative ($A_{ij} = 0, 
\pm1$) interactions are identical (blue dots) and coincide with the theoretical prediction (black curves) 
(Eq.~\beqref{QuantizationTheory}).
The results for the case $A_{ij} = \pm 1$ (red dots) shows a similar but different quantization 
behavior, following Eq.~\beqref{QuantizationTheoryNeg}.
(\textit{B}) 
Synchronization landscape $\sigma_{\min}$ as a function of both $m$ and $n$, computed from Eq.~\beqref{QuantizationTheory}.
Through simulated annealing, a network with initially high $\sigma$ (top red dot) evolves to a network 
satisfying Eq.~\beqref{eqn:ev} and $\sigma = \sigma_{\min}$ (bottom red dot).
(\textit{C}) The coupling cost $K$ as a function of $m$ and $n$ in units of $K_2$ (the cost for 
a network of two coupled units).  
The cost takes the lower values along the lines of discontinuity ($m = q_k$). 
}
\end{figure}

\ 
A surprising consequence of having these optimal synchronization properties 
is the quantization of the number of links in the networks.
We find, for example, that for binary interaction networks (i.e., $A_{ij}= 0,1$) satisfying condition~
\beqref{Optimal}, the number of links $m= \sum_i\sum_{j \neq i} A_{ij}$ is quantized to multiples of $n-1$.
That is,
\begin{equation}\label{quantization}
m = q_k := k(n-1), \quad \text{where $k = 1,2,\ldots, n$.}
\end{equation}
This follows from the identity $m = \sum_i L_{ii} = \sum_{i=2}^n \lambda_i = 
(n-1)\bar\lambda$, combined with the fact that 
condition~\beqref{Optimal} constrains the real eigenvalue $\bar\lambda$ further to be an integer for 
networks with integer-valued interactions (see Supporting Information, Section 1).
Consequently, any network with $m$ strictly between these quantized values must have $\sigma > 0$, 
and hence cannot be optimal.
What is then the minimum $\sigma$ for all such networks with a given $m$?  We denote this minimum 
by $\sigma = \sigma_{\min}(m)$.
Based on our analysis of all networks with $n \le 6$, we conjecture that the condition to achieve this 
minimum  
is that the Laplacian eigenvalues  (counting multiplicity) have the form
\begin{equation}\label{eqn:ev}
0, \stackrel{q_{k+1}-m}{\overbrace{k, \ldots, k}},  \stackrel{m - q_k}{\overbrace{k+1, \ldots, k
+1}},
\end{equation}
where $k$ is the integer satisfying $q_k \le m \le q_{k+1}$.
Note that, analogously to Eq.~\beqref{quantization} for $m = q_k$, this condition asserts that 
the eigenvalues are not only real but also integer for any $m$.
This leads to our prediction that
\begin{equation}\label{QuantizationTheory}
\sigma_{\min}(m) = \frac{1}{(n-1)d}\sqrt{(m - q_k)(q_{k+1} - m)},
\end{equation}
which is expected to be valid for binary interaction networks with arbitrary number of nodes and links.
Indeed, Fig.~\ref{Quantization}\textit{A} shows that for $10 \le n \le 12$, a simulated annealing 
procedure identified networks (blue dots) with $\sigma$ within $10^{-3}$ of (but not smaller than) $
\sigma_{\min}(m)$ predicted by Eq.~\beqref{QuantizationTheory} (black curve).
Starting from the (optimal) fully connected network [with $m = n(n-1)$] at the right end of the 
curve $\sigma_{\min}(m)$, any initial link deletion necessarily makes synchronization more difficult. 
Further deletions, however, can make it easier, bringing the networks back to optimal periodically as a 
function of $m$ at the cusp points and eventually leading to (optimal) directed spanning trees with $m 
= n-1$ (see Supporting Video).
The optimization of synchronization at these cusp points is analogous to the optimization of the 
dynamical range of excitable networks at the critical points of phase transitions \cite{Kinouchi:2006nx}.
Similar cusps are also observed for the Laplacian eigenvalues of a structurally perturbed optimal network as a function of the perturbation
(see Supporting Information, Section 2).
Note that, although the cost $K = \bar\eps_s \cdot m$ generally depends on $m$, it is actually 
independent of $m$ for optimal networks of given size $n$ because the synchronization 
threshold $\bar\eps_s  = \bar\eps_s(m)$ compensates for any change in $m$.

To reveal the intricate dependence of synchronization on the number of nodes and links, we now 
consider $\sigma_{\min}$ as a function of both $m$ and $n$ (Fig.~\ref{Quantization}\textit{B}) 
based on our prediction~\beqref{QuantizationTheory}.
Each point on this synchronization landscape represents \emph{all} networks with $\sigma = 
\sigma_{\min}$ for a given $m$ and $n$, and the number of such networks is expected to grow 
combinatorially with $n$ (see Supporting Information, Section 3).
All other networks with the same $m$ and $n$ are represented by points directly above that point.
The evolution of one such network by rewiring links under pressure to optimize synchronizability can be 
visualized as a vertical descent from a point above the landscape toward a point on the landscape that 
minimizes $\sigma$ (red arrow in Fig.~\ref{Quantization}\textit{B}).
The sharp valleys are observed along the lines $m = q_k$, and therefore correspond to the cusp points 
in Fig.~\ref{Quantization}\textit{A}.
Because two adjacent points on the landscape may include networks that do not have similar 
structures, it is surprising that one can actually move from one point to the next often by a simple 
structural modification (see Supporting Video).
Indeed, moving in the direction of the $m$-coordinate can be achieved by simply adding or removing a 
link that induces the smallest increase or largest decrease in $\sigma$.
Along the line $m = q_k$, 
an optimal network 
can be ``grown'' and kept 
optimal, by adding a node and connecting any $k$ existing nodes to the new node (see 
Supporting Information, Section 3).
The flexibility of choosing new links in this construction strongly suggests that optimal networks can 
have arbitrarily complex connectivity structures as the network size grows.
Another interesting landscape appears when we compute the cost $K$ as a function of 
$m$ and $n$ (Fig.~\ref{Quantization}\textit{C}) based on condition~\beqref{eqn:ev}.
In this landscape, optimal networks lie along the lines of discontinuity, resulting from the discontinuous 
change in $\bar\eps_s$ that occurs when $m$ changes from $q_k-1$ to $q_k$ and defines a saw-tooth function along the $m$-coordinate.
Note that for optimal networks, the cost $K$ is independent of $m$, as mentioned above, but 
increases linearly with $n$ and can be expressed as $K = K_2 (n-1)$, where $K_2$ is the 
coupling cost for a network of two coupled units.
A different but potentially related structure is the roughness of the time horizon considered in the 
synchronization of parallel processing units in distributed computing \cite{guclu:066115,G.Korniss01312003}.

While the presentation above focused on networks of identical units, the main results also hold true for 
networks of non-identical units.
Adopting the framework of Ref.~\cite{Sun:2009hc} and developing further analysis for networks of one-dimensional maps, we show in Supporting Information (Section 4) that
complete synchronization is possible even for non-identical units.
Moreover, we show that this is possible only for networks satisfying Eq.~\beqref{Optimal} in 
addition to the condition that the Laplacian matrix $L$ is diagonalizable.
Since any such networks must also exhibit the same quantization expressed in Eq.~\beqref{quantization}, we also expect cusps similar to those shown in Fig.~\ref{Quantization}.
For each quantized number of links $q_k$, we can show that there is a network that can 
synchronize completely despite the heterogeneity of the dynamical units.
Therefore, for both identical \emph{and} non-identical units, condition~\beqref{eqn:ev} can be used 
to 
systematically construct examples of suboptimal networks that can be made optimal by either adding 
or removing links.

\subsection*{Stabilizing effect of negative interactions.}\ 
Interestingly, the exact same quantization effect described above for binary interaction networks is also 
observed when we allow for negative interactions and interpret $m = \sum_i\sum_{j \neq i} A_{ij}$ as the 
\emph{net} number of links.
To see this, we use a generalization of the complement network, which we define for a given constant $
\alpha$ to be the network with adjacency matrix $A^c$ given by 
\begin{equation}\label{comp_trans}
A_{ij}^c := (\alpha - A_{ij})(1-\delta_{ij}).
\end{equation}
(This includes the special case $\alpha = 1$, which for undirected unweighted networks corresponds to 
the previously studied case of complement graphs \cite{MR2159259}.)
The transformation from a network to its complement 
maps $m$ to $\alpha n(n-1) - m$, $\lambda_i$ to $\alpha n - \lambda_i$, $\sigma$ to $\frac{m
\sigma}{\alpha n(n-1)-m}$, and thus an optimal network to another optimal network when $\alpha > 
\frac{m}{n(n-1)}$ (see {\it Materials and Methods}).
This also establishes a mapping between networks capable of complete synchronization of non-identical units, since the Laplacian matrix for such a network remains diagonalizable under the transformation (see Supporting Information, Section 4).
The condition on $\alpha$ avoids (nonsynchronizable) networks having eigenvalues with negative real 
part as an artifact of the transformation.
We argue that this generalized complement transformation is a powerful tool in analyzing networks 
with negative interactions because it reduces problems involving negative interactions to those 
involving only positive interactions when we choose $\alpha \ge \max A_{ij}$.
 
As an example of quantization with negative interactions, we consider the class of networks for which 
$A_{ij} = 0, \pm 1$ and assume that $\lambda_2,\ldots,\lambda_n$ have positive real parts to 
ensure that the network can synchronize.
Mapping these networks under the complement transformation with $\alpha = 1$, we obtain positive 
interaction networks with $A_{ij} = 0, 1, 2$.
Conjecture~\beqref{eqn:ev} applied to the resulting networks then leads to a prediction 
\emph{identical} to Eq.~\beqref{QuantizationTheory}, which we validated by simulated annealing for 
networks of size up to 12 (blue dots, Fig.~\ref{Quantization}\textit{A}). 
Thus, all the results discussed above based on Eq.~\beqref{QuantizationTheory}, including Eq.~
\beqref{quantization} and the shape of the synchronization landscape, are predicted to be valid even 
when negative interactions are allowed.
Whether the networks with $\sigma = \sigma_{\min}(m)$ actually do have negative interactions 
is not a priori clear because the synchronization landscape can be built entirely by using the subset of networks with only positive interactions.
Our simulated annealing shows, however, that many optimal networks (i.e., with $\sigma=0$) have 
negative interactions, as illustrated by an example in Fig.~\ref{ExampleOptimal}.
In addition, complete synchronization of non-identical units is possible in the presence of negative 
interactions (see Supporting Information, Section 4).
These networks provide clear and systematic examples of negative interactions that improve 
synchronization, as removing negative interactions from any such network would in 
general push $m$ off the quantized values ($m = q_k$) and make the network suboptimal.

The quantization of $m$ and the shape of the synchronization landscape, though they were identical 
for the two examples above, do depend critically on the constraints imposed on the interactions.
Consider, for example, the fully connected networks with interaction strengths $\pm 1$.
It can be shown that the implied constraint $A_{ij} \neq 0$ leads to a different quantization compared 
to Eq.~\beqref{QuantizationTheory}, which involves multiples of $2(n-1)$ rather than $n-1$,
\begin{equation}\label{QuantizationTheoryNeg}
\sigma_{\min}(m) = \frac{1}{(n-1)d}\sqrt{\bigl(m - q_{n-2(k+1)}\bigr)(q_{n - 2k} - m)}
\end{equation}
for $q_{n-2(k+1)} \le m \le q_{n - 2k}$ (red dots, Fig.~\ref{Quantization}\textit{A}). 
The synchronization landscape can be drastically different in some extreme cases.
On the one hand, within the widely studied class of undirected unweighted networks (corresponding to a 
stronger set of constraints, $A_{ij} = A_{ji}, A_{ij} = 0, 1$), no network with $m < n(n-1)$ satisfies 
the optimality condition~\beqref{Optimal} \cite{Donetti:2005kl,Nishikawa:2006kx}, which indicates 
that $\sigma_{\min} (m) > 0$ for all $m < n(n-1)$.
On the other hand, within the class of weighted networks corresponding to having no constraint on the 
interaction strengths, an optimal network can be constructed for any number of links $\ge n-1$ (e.g., the hierarchical 
networks in 
Ref.~\cite{Nishikawa:2006kx}), 
which implies that $\sigma_{\min}$ is identically zero in this particular case.

\begin{figure}[t]
\begin{center}
\includegraphics[width=3in]{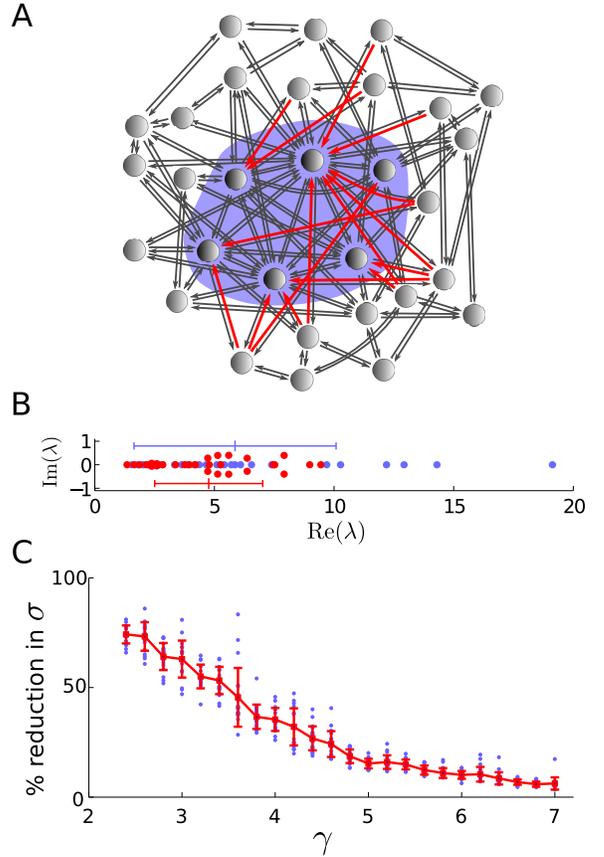}
\end{center}
\caption{\label{neg}
Improving synchronization with negative interactions.
(\textit{A}) Strength $-1$ is preferentially assigned to directional links (red arrows) connecting to high in-degree nodes 
 (those in the shaded area) in an initially undirected network.
The black links have strength $+1$. 
(\textit{B}) The initial spread of the Laplacian eigenvalues is significantly 
reduced as a result (from blue to red dots).
The error bars indicate the mean and one standard deviation of the corresponding distributions.
(\textit{C}) The mean and the standard deviation (red error bars) of the percentage reduction in $
\sigma$ over 20 realizations (blue dots) of random undirected scale-free networks with 1,000 nodes, 
minimum degree 5, and the scaling exponent $\gamma$.
The networks were generated by connecting nodes randomly according to a power-law degree distribution 
$P(\kappa) \sim \kappa^{-\gamma}$.
}
\end{figure}

To demonstrate that the synchronization enhancement by negative interactions goes much beyond the 
realm of optimal networks, we propose a simple algorithm for assigning
strength $-1$ to directional links in an arbitrary network with all link strengths initially equal to $+1$.
Our strategy is based on the observation that the in-degree distribution is a main factor determining 
synchronizability~\cite{Zhou:2006wb,Motter2004af,Motter:2005ub,1367-2630-9-6-182,Belykh:2005zr}, where the in-degree (or the total input 
strength) of node $i$ is defined to be $\sum_{j\neq i} A_{ij}$.
Since heterogeneity in the in-degree distribution tends to inhibit synchronization \cite{Zhou:2006wb,Motter2004af,Motter:2005ub,1367-2630-9-6-182,PhysRevLett.92.074103}, here we use negative interactions to 
compensate for positive interactions and to homogenize the in-degree distribution.
For a given network, we first choose randomly a node with the smallest in-degree and change the 
strength of each out-link of that node to $-1$, unless it makes the in-degree of the target node smaller 
than the mean in-degree of the original network.
We keep strength $+1$ for the other out-links, as well as all the in-links. 
Having treated this node, we repeat this process for the subnetwork of untreated nodes, considering links and hence degrees 
only within that subnetwork.
We continue this process until all nodes are treated.
Applying this algorithm to the network in Fig.~\ref{neg}\textit{A}, 
we see that the high in-degree nodes of the initial network (those in the shaded area) receive all of the 
compensatory negative interactions (red arrows), reducing $\sigma$ by nearly 35\% (Fig.~\ref{neg}
\textit{B}). 
The effectiveness of the method was further validated (Fig.~\ref{neg}\textit{C}) using random scale-free networks \cite{Barabasi:1999kc} generated by the standard configuration model \cite{PhysRevE.64.026118}, 
where we see more dramatic effect for more heterogeneous networks, reducing $\sigma$ by as much 
as 85\%.
The synchronization enhancement was indeed accompanied by the homogenization of the in-degree distribution (see Supporting 
Information, Section 5).
The use of negative directional interactions in our algorithm suggests that the enhancement is partly due to link directionality (see Ref.~\cite{PhysRevLett.103.228702} for an enhancement method purely based on link directionality), which generally plays an important role 
in synchronization (see for example Refs.~\cite{restrepo:015107} and \cite{belykh:015102}).
However, negative strength of interactions alone can also produce similar enhancement in random scale-free 
networks when they are assigned to bidirectional links between hubs (see Supporting Information, 
Section 6).

\section*{Conclusions}
Even though negative interactions and link removals by themselves tend to destabilize synchronous states, we 
have shown that they can compensate for other instabilities, such as those resulting from a 
``forbidden" number of interactions or from a heterogeneous in-degree distribution. 
This establishes an unexpected relation between network synchronization and recent work on metabolic 
networks, where locally deleterious perturbations have been found to 
generate similar compensatory effects that are globally beneficial in the presence of other deleterious 
perturbations \cite{Motter:2008rm, Motter:2010}.
For example, the removal of a metabolic reaction --- or, equivalently, of its enzyme-coding gene(s) --- can often be partially or totally compensated by the targeted removal of a second metabolic reaction. 
That is, the inactivation of specific metabolic reactions can improve the performance of defective or 
sub-optimally operating metabolic networks, and in some cases it can even rescue cells that would be 
nonviable otherwise \cite{Motter:2008rm}. Other research has shown that similar reaction inactivation 
occurs spontaneously for cells evolved to optimize specific metabolic objectives \cite{Nishikawa:2008kx}. 
These apparently disparate examples share the common property that the collective dynamics is 
controlled, and in fact enhanced, by {\it constraining} rather than augmenting the underlying 
network. 
Such network-based control, including the conditionally positive impact of otherwise detrimental 
interactions and constraints, is most likely not unique to these contexts.

In neuronal networks, for example, inhibitory interactions have been predicted to facilitate synchronous 
bursting \cite{Bragin:1995uq,Whittington-MA:1995yq,Ylinen:1995kx,Belykh:2008zf,Vreeswijk:1994sm,Wang:1996dp}. 
Even more compelling, it has been established for both animal \cite{Bourgeois:1993iz} and human 
\cite{PR.:1979kx,Huttenlocher:1997vn} cerebral cortices that the density of synapses first increases and then 
decreases during early brain development, suggesting a positive role played by link removal also in 
neuronal networks. 
More generally, we expect that the eigenvalue spectrum analysis that lead to our conclusions will help 
investigate analogous behavior in yet other classes of network processes governed by spectral 
properties, such as epidemic spreading \cite{PhysRevE.66.047104} and cascading failures 
\cite{Motter:2004pi,Simonsen:2008rc}. 

Taken together, the highly structured characteristics revealed by the synchronization landscape explain 
why the characterization of the network properties that govern synchronization has been so elusive. 
Numerous previous studies performed under comparable conditions have sometimes led to apparently conflicting conclusions about the role played by specific network structural properties such 
as randomness, betweenness centrality, and degree distribution \cite{Arenas:2008yq}. 
Our results indicate that at least part of these disagreements can be attributed to the sensitive 
dependence of the synchronization properties on the specific combination of nodes and links, as clearly 
illustrated by the non-monotonic, periodic structure of cusps exhibited by the synchronization 
landscape. 
We suggest that insights provided by these findings will illuminate the design principles and evolution mechanisms of both natural and engineered networks in which synchronization is functionally important~\cite{Kiss:2007kx}.

\begin{materials}

\subsection{Synchronization stability analysis}
The stability analysis can be carried out using a master stability approach based on linearizing Eq.~
\beqref{MainEqn} around synchronous states \cite{Pecora:1998zp}.
We apply a coordinate transformation to reduce the Laplacian matrix $L$ to its Jordan canonical form 
and decompose the variational equations into components along the corresponding eigenspaces 
\cite{Nishikawa:2006fk,Nishikawa:2006kx}.  
This leads to a master stability equation,
\begin{equation}\label{MSE}
\dot{\mathbf{y}} = [D\mathbf{F}(\mathbf{s}) - a D\mathbf{H}(\mathbf{s})] \mathbf{y},
\end{equation}
where 
$a = \bar\eps\lambda_i$
for the eigenspace corresponding to $\lambda_i$.
The stability function $\Lambda(a)$ is the maximum Lyapunov exponent for the solution $\mathbf{y}
(t) = \mathbf{0}$ of Eq.~\beqref{MSE}.
The $A_{ij}$ are not required to be nonnegative and, since the derivation of $\Lambda(a)$ is based on a Jordan form, $L$ is not required to be diagonalizable either~\cite{Nishikawa:2006fk,Nishikawa:2006kx}.
Geometrically, the stability condition is that all the (possibly complex) numbers 
$\bar\eps\lambda_2, \ldots, \bar\eps\lambda_n$ 
lie in the region $\{ a \in \mathbb{C} : \Lambda(a) < 0 \}$ of the complex plane.
Since $\Lambda(a)$ is generally positive on the left side of the imaginary axis, we consider only the 
networks whose Laplacian eigenvalues have nonnegative real part, which ensures that complete 
synchronization is possible.

\subsection{Dynamical robustness of optimal networks}
For a given stability function $\Lambda(a)$, a network satisfying condition~\beqref{Optimal} has the 
maximum rate of exponential convergence to synchronization among all networks. 
This is so under the assumption that there is a physical limit $M$ to the coupling strength of individual 
links, $\bar\varepsilon A_{ij} \le M$, and that the stability function $\Lambda(a)$ is continuous and 
monotonically increasing with respect to the distance from the real axis, which appears to hold true for all 
known examples of $\Lambda(a)$ in the literature~\cite{huang:036204}.
From the limit on coupling strengths, it follows that the real part of $\bar\varepsilon \lambda_i$ is 
bounded by a constant $Mn(n-1)$, which,
combined with the assumption on $\Lambda(a)$, implies that the exponential rate of convergence is 
completely determined by the value of $\Lambda(a)$ in the interval $[0, Mn(n-1)]$ on the real axis.
In this interval $\Lambda(a)$ has a global minimum at some $a^*$, and thus $r^* := -\Lambda(a^*) 
> 0$ is the maximum possible rate of exponential convergence.
If the network satisfies condition~\beqref{Optimal}, 
all perturbation eigenmodes converge at the maximum rate $r^*$ when we choose 
$\bar\eps = \frac{a^*}{\bar\lambda}$.
In contrast, if the network violates condition~\beqref{Optimal}, there must be at least one eigenvalue $
\lambda_i$ such that 
$\bar\varepsilon\lambda_i \neq a^*$ [excluding the exceptional situation where multiple values of $
\bar\varepsilon\lambda_i$ fall precisely on multiple global minima of $\Lambda(a)$],
resulting in an eigenmode that converges at a slower rate 
$r := -\Lambda(\bar\varepsilon\lambda_i) < r^*$.
Therefore, although optimal networks may suffer from initially slower convergence to synchronization 
that is polynomial in time \cite{Nishikawa:2006fk,Nishikawa:2006kx}, the long-term convergence rate 
is dominated by $r^*$ and is faster than for any other network.
Indeed, the deviation from the synchronous state can be written as $P(t)e^{-r^* t}$ for an optimal 
network and as $Q(t) e^{-rt}$ for a suboptimal network, where $P(t)$ and $Q(t)$ are polynomials. 
The ratio between the deviations in the two cases is then
\begin{equation}
\frac{P(t)}{Q(t)} e^{-(r^*-r) t} \to 0 \quad \text{ as $t \to \infty$}
\end{equation}
and, in particular, is less than $1$ for sufficiently large $t$, implying that the deviation will eventually 
become smaller for the optimal network.

\subsection{Laplacian spectrum of generalized complements}
We show  
that if the eigenvalues of $L$ are $0, \lambda_2, \ldots, \lambda_n$ (counting multiplicity), then the eigenvalues of the Laplacian matrix $L^c$ of the generalized complement (defined by Eq.~\beqref{comp_trans}) are $0, n\alpha - \lambda_2, \ldots, n\alpha - \lambda_n$.
This result follows from the relation
\begin{equation}\label{eq.polyn}
\mu(L^c, x) = (-1)^{n+1} \frac{x}{n\alpha - x} \, \mu(L, n\alpha - x),
\end{equation}
where $\mu(L, x) = \det(L - x I)$ is the characteristic polynomial of the matrix $L$ and $I$ is the $n \times n$ identity matrix.
We derive this relation by following the strategy of the proof of Lemma 2.3 in Ref.~\cite{Mohar:1990ul} for undirected networks with nonnegative link strengths, to now consider directional and weighted links, possibly with negative strengths.
From the definition of the
complement 
transformation, we can write $L + L^c = n\alpha I - \alpha J$, where $J$ is the $n \times n$ matrix with every entry equal to one.
Using this and well-known properties of the determinant, we have
\begin{align}
\mu(L^c, x) &= \det(L^c - x I) \nonumber \\
&= \det(n\alpha I - \alpha J - L - x I) \nonumber\\
&= (-1)^n \det \bigl[L^T + \alpha J - (n\alpha - x) I\bigr] \nonumber\\
&= (-1)^n \mu(L^T + \alpha J, n\alpha - x),
\end{align}
where $L^T$ denotes the transpose of $L$.
Eq. \beqref{eq.polyn} 
then follows from the fact that $\mu(L^T + \alpha J, z)/(n\alpha-z) = -\mu(L^T, z) / z $ whenever $L$ has row sums equal to zero (which is the case, since $L$ is the Laplacian matrix of a network).
To prove this fact, we will use  
elementary row operations, which do  
not change the determinants. 
First, we replace the first row of the matrix $L^T + \alpha J - z I$ by the sum of all rows, making each entry in this row $n\alpha-z$.
Next, we subtract this row multiplied by $\alpha/(n\alpha-z)$ (which makes it a row with all entries equal to $\alpha$) from the remaining rows, canceling out the contribution from the term $\alpha J$ in these rows.
We denote the resulting matrix by $M_1$.
Finally, we take the matrix $L^T - z I$ and replace the first row by the sum of all rows.  This results in a matrix $M_2$ with exactly the same entries as $M_1$ except for the first row.  For this matrix this is a row with all entries equal to $-z$ rather than $n\alpha-z$. 
Dividing the first rows of $M_1$ and $M_2$  
by $n\alpha-z$ and $-z$, respectively, which will scale the determinants accordingly, we see that 
\begin{equation}
\begin{split}
\frac{\mu(L^T + \alpha J, z)}{n\alpha-z} &=
\frac{\det(L^T + \alpha J - z I)}{n\alpha-z} = \frac{\det(M_1)}{n\alpha-z} \\&= \frac{\det(M_2)}{-z} = \frac{\det(L^T - z I)}{-z}
= -\frac{\mu(L^T, z)}{z}.
\end{split}
\end{equation}
As an immediate consequence of this result, the
normalized standard deviation $\sigma$ changes to $\frac{m\sigma}{\alpha n(n-1) - m}$ 
under the complement transformation, while the total link strength
$m$ is mapped to $\alpha n(n-1) - m$.
In particular, this implies that the complement of any optimal network is also optimal if $\alpha > \frac{m}{n(n-1)}$.

\end{materials}

\begin{acknowledgments}
The authors thank 
Marian Anghel for discussions on the power-grid network and
Sara Solla for providing important references on the role of inhibitory neurons.
This work was supported by NSF under Grant No. DMS-0709212.
\end{acknowledgments}


\newpage


\newpage

\newpage



\newpage

\setcounter{figure}{0}
\renewcommand{\thefigure}{S\arabic{figure}}

\setcounter{equation}{0}
\renewcommand{\theequation}{S\arabic{equation}}

\section{\huge Supporting Information}
\medskip

\section{1. Laplacian spectrum of optimal networks}\label{spec_opt}

Consider a network whose (possibly directional) links have integer strengths, and denote its Laplacian matrix by $L$. 
Here we show that if the network is optimal, i.e., the non-identically zero eigenvalues of $L$ assume a common value $\bar\lambda$ (Eq.~\textbf{3} in the main text),
then $\bar\lambda$ must be an integer. This result will be valid even when the links are allowed to
have negative integer strength.

The characteristic polynomial of $L$ can be written as
\begin{align}
\det(L - x I) &= - x(\bar\lambda-x)^{n-1}\nonumber\\
&= - \bar\lambda^{n-1} x + \cdots + (-1)^n x^n,
\end{align}
where $n$ is the number of nodes and $I$ is the $n \times n$ identity matrix.
Since $L$ has integer entries, all the coefficients of the characteristic polynomial are integers, and hence $\bar\lambda^{n-1}$ in the first term above is an integer.
Denote this integer by $k$.
Using $m$ to denote the sum of all link strengths in the network, we have $m = \sum_i L_{ii} = (n-1)\bar\lambda$, and hence $k = [m/(n-1)]^{n-1}$.
Writing $m/(n-1) = s/t$, where integers $s$ and $t$ do not have common factors, we obtain $kt^{n-1} = s^{n-1}$.
Suppose $p$ is a prime factor of $k$.
Then $p$ is also a factor of $s^{n-1}$, so in fact $p$ is a factor of $s$.
This implies that $p^{n-1}$ is a factor of $kt^{n-1} = s^{n-1}$.
Since $s$ and $t$ cannot have a common factor, $p^{n-1}$ must be a factor of $k$.
Thus, any prime factor of $k$ must actually appear with multiplicity $n-1$, and hence we can write $k = q^{n-1}$ where $q$ is an integer.
Therefore, $\bar\lambda^{n-1} = k = q^{n-1}$, and since $\bar\lambda$ is real, we have $\bar\lambda = q$, an integer.

\section{2. Perturbation of Laplacian eigenvalues}

Suppose that the Laplacian matrix $L_0$ of a given network of $n$ nodes has an eigenvalue $\lambda_0 \neq 0$ with multiplicity $k \le n-1$.
Consider a perturbation of the network structure in the form $L = L_0 + \delta L_1$, where $\delta$ is a small parameter and $L_1$ is any fixed Laplacian matrix 
representing the perturbed links.
We do not need to assume that $L_0$ and $L_1$ have nonnegative entries, making our result valid even in the presence of negative interactions.
Denote the characteristic polynomial of $L$ by $f(x, \delta) := \det(L - x I)$, where $I$ is the $n \times n$ identity matrix.
Since $\lambda_0$ is an eigenvalue of $L_0$ with multiplicity $k$, we have
\begin{equation}\label{char_poly}
f(x, 0) = (x - \lambda_0)^k g(x),
\end{equation}
where $g$ is a polynomial satisfying $g(\lambda_0) \neq 0$.
Denote by $\lambda = \lambda(\delta)$ an eigenvalue of $L$ that approaches $\lambda_0$ as $\delta \to 0$. 
Here we show that the change $\Delta\lambda := \lambda - \lambda_0$ of the eigenvalue induced by the perturbation scales as 
\begin{equation}         
\Delta\lambda \sim \delta^{1/k}
\label{eq.scaling}
\end{equation}
if the derivative of the characteristic polynomial with respect to the perturbation parameter evaluated at $\delta =0$
is nonzero:
\begin{equation}\label{pertub_cond2}
\left.\frac{\del f}{\del \delta}\right|_{\begin{subarray}{l}x = \lambda_0\\ \delta=0 \end{subarray}} \neq 0. 
\end{equation}
Through the Jacobi's formula for the derivative of determinants, this condition can be expressed as
\begin{equation}\label{pertub_cond}
\tr \bigl[ (L_0 - \lambda_0 I)^{k-1}  g(L_0) L_1 \bigr] \neq 0,
\end{equation}
where $\tr(A)$ denotes the trace of matrix $A$. 
We expect this condition to be satisfied for most networks 
and perturbations. 
For the optimal networks satisfying Eq.~\textbf{3} in the main text, it can be shown that Eq.~\beqref{pertub_cond} 
is violated if $L_0$ is diagonalizable, but $L_0$ is actually known to be nondiagonalizable for 
the majority of these optimal networks~\cite{Nishikawa:2006kx2}. 
The scaling \beqref{eq.scaling} 
shows that, for a fixed $\delta$, the more degenerate the eigenvalue (larger $k$), the larger the effect of the perturbation on that eigenvalue. 
In particular, if the original network is optimal with $\bar\lambda$ having the maximum possible multiplicity $n-1$, the effect of perturbation is the largest.
This, however, is so because the optimal networks have significantly smaller $\sigma$ than suboptimal networks (even those with just one more or one less link), and therefore the perturbations of the optimal networks would still be more synchronizable in general than most suboptimal networks.

From Eq.~\beqref{char_poly} it follows that
\begin{equation}\label{eq_app_2}
\left.\frac{\del f}{\del x}\right|_{\begin{subarray}{l}x = \lambda_0\\ \delta=0 \end{subarray}}
= \left.\frac{\del^2 f}{\del x^2}\right|_{\begin{subarray}{l}x = \lambda_0\\ \delta=0 \end{subarray}} 
= \cdots 
= \left.\frac{\del^{k-1} f}{\del x^{k-1}}\right|_{\begin{subarray}{l}x = \lambda_0\\ \delta=0 \end{subarray}}
= 0,
\end{equation}
but 
\begin{equation}\label{eq_app_3}
\left.\frac{\del^k f}{\del x^k}\right|_{\begin{subarray}{l}x = \lambda_0\\ \delta=0 \end{subarray}} = k! \cdot g(\lambda_0) \neq 0.
\end{equation}
Using this to expand $f(x,\delta)$ around $x = \lambda_0$ and $\delta = 0$ up to the $k$th order terms, and setting $x = \lambda$, we obtain 
\begin{equation}\label{eq8}
\frac{f(\lambda,\delta)}{\delta} 
= \left.\frac{\del f}{ \del \delta}\right|_{\begin{subarray}{l}x = \lambda_0\\ \delta=0 \end{subarray}} 
+ \frac{1}{k!} \left.\frac{\del^k f}{\del x^k}\right|_{\begin{subarray}{l}x = \lambda_0\\ \delta=0 \end{subarray}}
\cdot \frac{(\lambda-\lambda_0)^k}{\delta} + O(\delta),
\end{equation}
where $O(\delta)$ includes all higher order terms. 
From the characteristic equation $f(\lambda,\delta) = \det(L - \lambda I) = 0$, the left hand side of Eq.~\beqref{eq8} is zero, so  
taking the limit $\delta \to 0$ leads to 
\begin{equation}
\lim_{\delta \to 0} \frac{(\Delta \lambda)^k}{\delta}
= - \frac{1}{g(\lambda_0)} 
\left.\frac{\del f}{\del \delta}\right|_{\begin{subarray}{l}x = \lambda_0\\ \delta=0 \end{subarray}},
\end{equation}
which implies the scaling \beqref{eq.scaling} when condition~\beqref{pertub_cond2} is satisfied.

\section{3. Complexity of optimal networks}

Here we first describe a systematic method for increasing the number of nodes $n$ in an optimal binary interaction network ($A_{ij} = 0,1$), while keeping the network optimal.
Given an optimal network with $\bar\lambda = k$ (which must be an integer by the result in Section~\ref{spec_opt} above), we construct a new network by adding a new node and connecting any $k$ existing nodes to the new node.
As a result, the Laplacian matrix has the form
\begin{equation}
L = \left(\begin{array}{ccc|c}
& & & 0 \\
& L_0 & & \vdots \\
& & & 0 \\
\hline
u_1 & \cdots & u_n & k
\end{array}\right),
\end{equation}
where $L_0$ is the Laplacian matrix of the original network, each $u_i$ is either $0$ or $-1$, and $u_1 + \cdots + u_n = - k$.
Since $L$ is a block triangular matrix, its eigenvalue spectrum consists of the eigenvalues of $L_0$, which are $0, k, \ldots, k$, and an additional $k$, which comes from the last diagonal element.
Thus, the new network is optimal with $\bar\lambda = k$.

We can argue that the number of optimal binary interaction networks grows combinatorially with $n$.
To this end, we first consider $C(n)$, the number of distinct Laplacian matrices corresponding to optimal networks with $n$ nodes.
For each optimal network with $n$ nodes and $\bar\lambda = k$, the above construction gives ${n \choose k}$ different Laplacian matrices corresponding to optimal networks with $n+1$ nodes.
Using the bound ${n \choose k} \ge n$, which is valid for $k = 1, \ldots, n-1$, we see that $C(n+1) \ge n\cdot C(n)$, which implies $C(n) \ge (n-1)!$ and gives a combinatorially growing lower bound for $C(n)$.
Since two different Laplacian matrices may represent isomorphically equivalent networks, $C(n)$ is an overestimate of the number of optimal networks with $n$ nodes.
However, given the gross underestimate coming from ${n \choose k} \ge n$ and the fact that we used only one out of potentially many possible schemes for adding a node while keeping the network optimal, we expect that the number of optimal binary interaction networks with $n$ nodes also grows combinatorially with $n$.

\section{4. Optimality for networks of heterogeneous units}

Consider a network of coupled non-identical units whose dynamics is governed by
\begin{align}\label{HeteroEqn}
x_i(t+1) = F(x_i(t),\mu_i) + \bar\varepsilon \sum_{j=1}^n A_{ij}[H(x_j(t), \mu_j) - H(x_i(t), \mu_i)], 
\end{align}
where $t$ represents the discrete time and $\bar\varepsilon = \varepsilon/d$ is the global coupling strength normalized by the average coupling strength per node, $d = \frac{1}{n} \sum_i\sum_{j\neq i} A_{ij}$.
The dynamics of unit $i$ follows $x_i(t+1) = F(x_i(t),\mu_i)$ in the absence of coupling with other units and is assumed to be one-dimensional for simplicity. 
Variation in the parameter $\mu_i$ represents the dynamical heterogeneity of the network, which we measure by the standard deviation $\sigma_\mu$ defined by $\sigma_\mu^2 := \frac{1}{n} \sum_i \delta\mu_i^2$, where $\delta\mu_i := \mu_i - \bar\mu$ and $\bar\mu := \frac{1}{n}\sum_i \mu_i$.
Here we choose the signal function to be $H(x,\mu) = F(x,\mu)$, which leads to a natural generalization of coupled map lattices~\cite{kaneko:279} to arbitrary coupling topology.
For example, for the one-dimensional periodic lattice in which each unit is coupled only to its two nearest neighbors with unit strength, system~\beqref{HeteroEqn} reduces to the well-studied system
$x_i(t+1) = (1-\varepsilon) F(x_i(t),\mu_i) + \frac{\varepsilon}{2} F(x_{i-1}(t), \mu_{i-1})
 + \frac{\varepsilon}{2} F(x_{i+1}(t), \mu_{i+1})$.

We consider a nearly synchronous state~\cite{Sun:2009hc2}, in which the deviation of the states of individual units around their average is small, i.e., $\delta x_i(t) := x_i - \bar{x}(t)$ is small, where $\bar{x}(t) := \frac{1}{n} \sum_i x_i(t)$.
Following the strategy used in Ref.~\cite{Sun:2009hc2} for continuous-time systems, we linearize Eq.~\beqref{HeteroEqn} around the average state $\bar{x}(t)$ and the average parameter $\bar\mu$ to obtain the variational equation in the vector form
\begin{equation}\label{eqn:variational}
\delta\mathbf{x}(t+1) = (I - \bar\varepsilon \tilde{L}) (a_t \delta\mathbf{x}(t) + b_t \delta\boldsymbol\mu),
\end{equation}
where $\delta\mathbf{x}(t) = (\delta x_1(t), \ldots, \delta x_n(t))^T$ is the state deviation vector, $\delta\boldsymbol{\mu} = (\delta\mu_1, \ldots, \delta\mu_n)^T$ is the parameter variation vector, and $I$ is the $n \times n$ identity matrix.  Matrix $\tilde{L}$ is the modified Laplacian matrix defined by $\tilde{L}_{ij} = L_{ij} - \frac{1}{n}\sum_k L_{kj}$~\cite{Sun:2009hc2}, and we denote $[a_t, b_t] := \bigl[\frac{\del F}{\del x}(\bar{x}(t), \bar\mu), \frac{\del F}{\del \mu}(\bar{x}(t), \bar\mu)\bigr]$.
As a result of the linearization, the deviation $\delta\mathbf{x}(t)$ can possibly diverge as $t\to\infty$ even when the state space for the network dynamics is bounded.
Notice that $(1,\ldots,1)^T$ is an eigenvector of the matrix $I - \bar\varepsilon\tilde{L}$ associated with eigenvalue one.
The component of the linearized dynamics parallel to this vector is irrelevant for synchronization stability, since $\delta\mathbf{x}(t)$ by definition does not have this component.
We thus remove this component, keeping all other components unchanged, by 
replacing $I$ in Eq.~\beqref{eqn:variational} with $\tilde{L}^*$ defined by $\tilde{L}^*_{ij} = \delta_{ij} - 1/n$, which leads to
\begin{equation}\label{eqn:variational2}
\delta\mathbf{x}(t+1) = 
(\tilde{L}^* -\bar\varepsilon \tilde{L}) (a_t \delta\mathbf{x}(t) + b_t \delta\boldsymbol\mu).
\end{equation}
Any component along $(1,\ldots,1)^T$ will 
immediately vanish under multiplication of the matrix $\tilde{L}^* -\bar\varepsilon \tilde{L}$, whose properties govern the evolution of synchronization error.

As a measure of synchronization error, we use the standard deviation $\sigma_x(t)$ defined by $\sigma^2_x (t) := \frac{1}{n} \sum_i \delta x^2_i(t)$.
For a fixed $\sigma_\mu$ and $\bar\varepsilon$, we define the maximum asymptotic synchronization error to be
\begin{equation}\label{eqn:omega_eps}
\Omega_{\bar\varepsilon}(L) := \max_{\{\mu_i\}}\,\limsup_{t\to\infty} \sigma_x(t),
\end{equation}
where the maximum is taken over all possible combinations of $\mu_i$ for the given $\sigma_\mu$.
We can explicitly compute $\Omega_{\bar\varepsilon}(L)$ by iterating Eq.~\beqref{eqn:variational2}, which leads to
\begin{multline}\label{eqn:omega_eps2}
\Omega_{\bar\varepsilon}(L) =\sigma_\mu \tilde\Omega(\bar\varepsilon \tilde{L}), \quad \text{where } \\
\tilde\Omega(X) :=
\limsup_{T\to\infty}\Biggl\lVert\sum_{t=1}^T \left(\prod_{k=1}^{t-1}a_{k+T-t}\right)
b_{T-t}(\tilde{L}^* -X)^t\Biggr\rVert,
\end{multline}
where $\norm{\cdot}$ denotes the spectral norm for matrices.
Notice that the synchronization error is a linear function of $\sigma_\mu$, which is a consequence of using Eq.~\beqref{eqn:variational2}.
The function $\tilde\Omega(X)$, whose argument is a matrix $X$ encoding the network structure, can be interpreted as a master synchronization error function since its functional form is determined entirely by the map $F(x,\mu)$ and the averaged trajectory $\bar{x}(t)$, and is independent of the network structure.
A sufficient condition for $\tilde\Omega(X)$ to be finite is $\rho(\tilde{L}^* -X) < e^{-\nu}$, where $\rho(\cdot)$ denotes the spectral radius of matrices, or equivalently, the maximum of the absolute values of the eigenvalues.
Here $\nu$ is the Lyapunov exponent of the averaged-parameter map $F(x,\bar\mu)$ along the average trajectory $\bar{x}(t)$, i.e., $\nu := {\displaystyle\lim_{T\to\infty}}\frac{1}{T}\sum_{t=0}^{T-1}\ln\bigl\lvert\frac{\del f}{\del x}(\bar{x}(t),\bar\mu)\bigr\rvert$.
If $X = \bar\varepsilon \tilde{L}$, this condition reduces to the stability condition for complete synchronization of the corresponding identical units, namely, $\Lambda(\bar\varepsilon\lambda_i) < 0$ for $i = 2,\ldots,n$, where the stability function in this case is $\Lambda(a) = \nu + \ln|1-a|$.
For example, if $F(x,\mu) = 2x+\mu \mod 1$, the sum in Eq.~\beqref{eqn:omega_eps2} converges to $\tilde\Omega(X) = \norm{(\tilde{L}^*-X)(2X-2\tilde{L}^*+I)^{-1}}$ when $\rho(\tilde{L}^* -X) < 1/2$.

\begin{figure}[t]
\begin{center}
\includegraphics[width=3.4in]{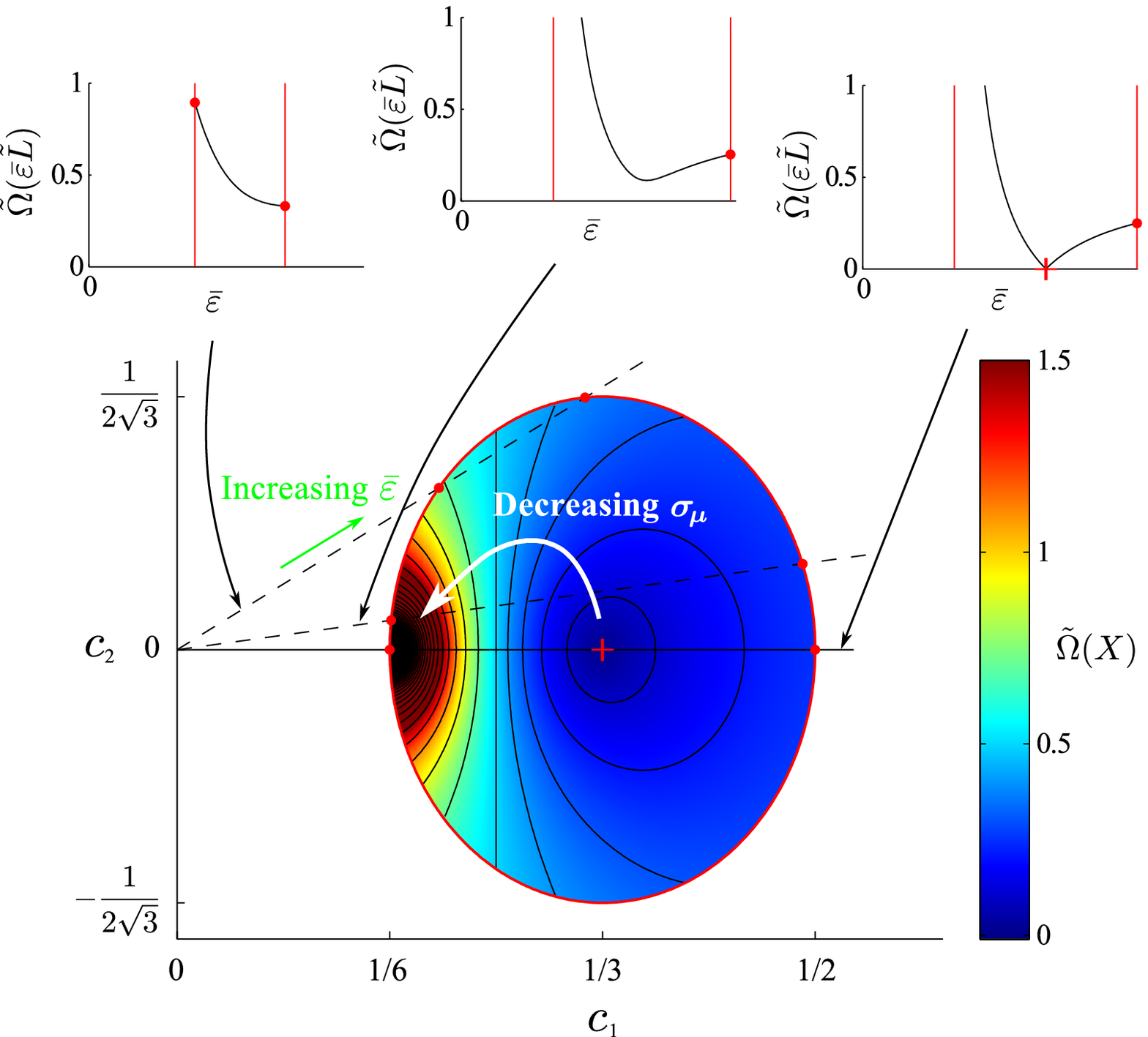}\rule{10mm}{0in}
\end{center}
\caption{\label{figS1}
Master synchronization error function $\tilde\Omega(X)$ defined on the space of matrices.
The color indicates the values of $\tilde\Omega(X)$ with $X = \tilde{L}$ for the two-parameter family of networks given by $L = c_1 L_{K_3} + c_2 (L_{C_{3,1}} - L_{C_{3,2}})$ and the map $F(x,\mu) = 2x + \mu \mod 1$.
Shown in black are the level curves $\tilde\Omega(X) = \frac{E_\text{tol}}{\sigma_\mu}$.
For fixed error tolerance $E_\text{tol}$ and 
heterogeneity $\sigma_\mu$, the region defined by $\tilde\Omega(X) \le \frac{E_\text{tol}}{\sigma_\mu}$ represents the set of networks for which synchronization error is within $E_\text{tol}$.
The red closed curve indicates the edge of synchronization stability defined by $\rho(\tilde{L}^* - X) < e^{-\nu}$.
Each dashed line represents the path $X = \bar\varepsilon\tilde{L}$ for a fixed $\tilde{L}$, which indicates the locations of synchronization transition (red dots) and, more generally, how synchronization error changes as $\bar\varepsilon$ is varied (insets).
The same representation for the path along the $c_1$-axis shows that $\tilde\Omega(X)=0$ when $X = \tilde{L}^*$, corresponding to the point $(c_1,c_2)=(1/3,0)$ for this example (indicated by a red plus symbol).
}
\end{figure}

\begin{figure}[t]
\begin{center}
\includegraphics[width=3.4in]{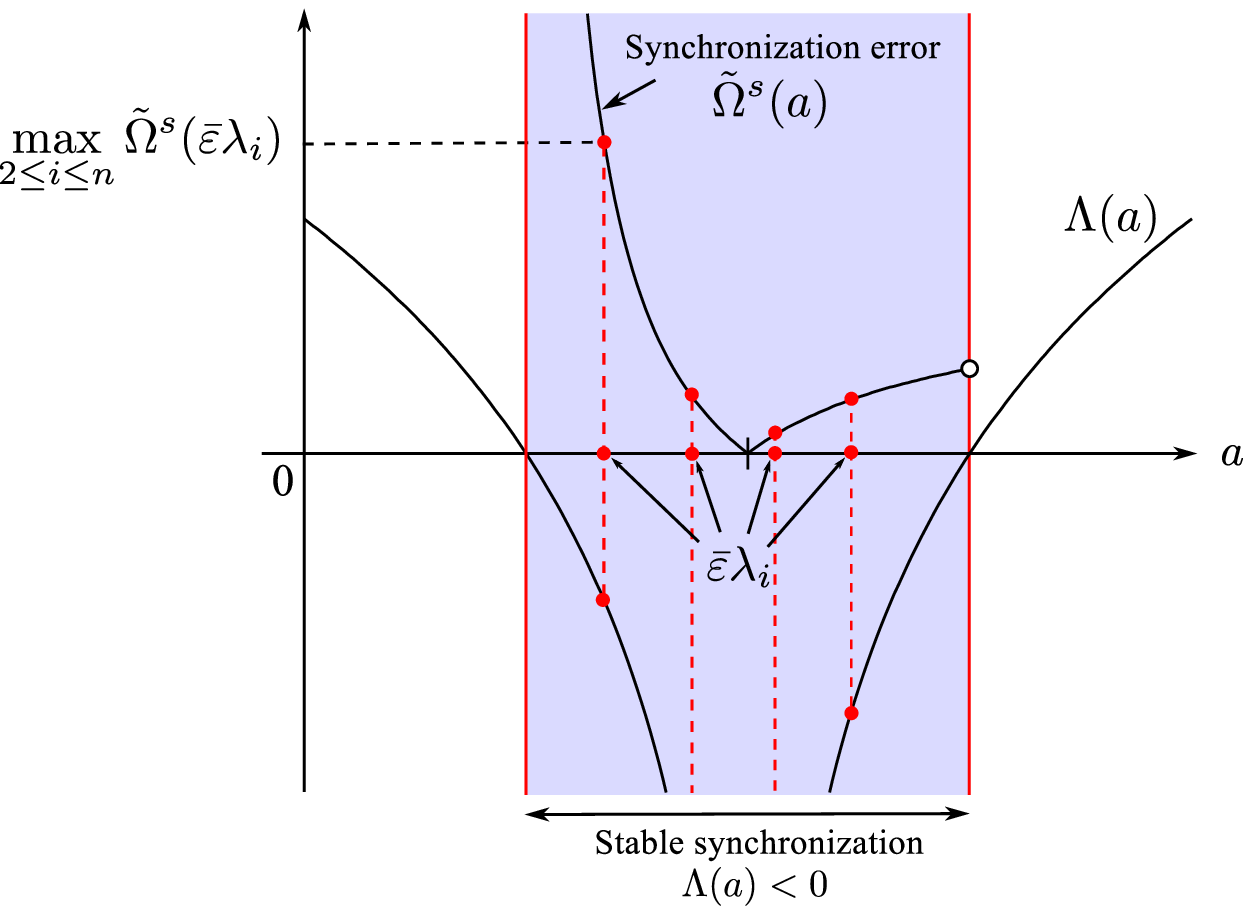}\rule{10mm}{0in}
\end{center}
\caption{\label{figS2}
Synchronization error function for undirected networks.
The condition for the stability of synchronization is that the numbers $\bar\varepsilon\lambda_2, \ldots, \bar\varepsilon\lambda_n$ (the red dots on the $a$-axis) all lie in the interval on which $\Lambda(a) < 0$.
When this condition is satisfied, the synchronization error is determined by the highest point among the corresponding points on the curve $\tilde\Omega^s(a)$ (the red dots on the top curve).
For illustration we used the curves for $F(x,\mu) = 2x + \mu \mod 1$, which are $\tilde\Omega^s(a) = \bigl\lvert\frac{1-a}{2a-1}\bigr\rvert$ and $\Lambda(a) = \ln 2 + \ln|1-a|$.
}
\end{figure}

The set of networks with a given synchronization error tolerance $E_\text{tol}$ defined by $\Omega_{\bar\varepsilon}(L) \le E_\text{tol}$ is represented by the region $\bigl\{ X: \tilde\Omega(X) \le \frac{E_\text{tol}}{\sigma_\mu} \bigr\}$ in the space of matrices.
In Fig.~\ref{figS1} we illustrate this using the two-parameter family of networks defined by
$L = c_1 L_{K_3} + c_2 (L_{C_{3,1}} - L_{C_{3,2}})$,
where $L_{K_3}$, $L_{C_{3,1}}$, and $L_{C_{3,2}}$ are the Laplacian matrices of the fully connected network of 3 nodes and the two types of 3-cycles:
\begin{align}
L_{K_3} &= \begin{pmatrix} 
2 & -1 & -1 \\
-1 & 2 & -1 \\
-1 & -1 & 2
\end{pmatrix},\quad
L_{C_{3,1}} = \begin{pmatrix} 
1 & 0 & -1 \\
-1 & 1 & 0 \\
0 & -1 & 1
\end{pmatrix},\nonumber\\ 
L_{C_{3,2}}&= \begin{pmatrix} 
1 & -1 & 0 \\
0 & 1 & -1 \\
-1 & 0 & 1
\end{pmatrix}.
\end{align}
For this family of networks, we can show that for $X = \tilde{L}$, we have
\begin{equation}
\tilde\Omega(\tilde{L}) = \limsup_{T\to\infty} \Biggl\lvert\sum_{t=1}^T \left(\prod_{k=1}^{t-1}a_{k+T-t}\right) b_{T-t} (1-\lambda)^t \Biggr\rvert,
\end{equation}
where $\lambda = 3c_1 + i\sqrt{3}c_2$.
In particular, we have $\tilde\Omega(\tilde{L}) = \bigl\lvert\frac{1-\lambda}{2\lambda-1}\bigr\rvert$ if $F(x,\mu) = 2x + \mu \mod 1$, and this is used as an illustrative example in Fig.~\ref{figS1}.
As $\sigma_\mu$ approaches zero and the dynamical units become less heterogeneous, the region $\bigl\{ X: \tilde\Omega(X) \le \frac{E_\text{tol}}{\sigma_\mu} \bigr\}$ increases in size and approaches the region of stable synchronization given by $\rho(\tilde{L}^*-X) < e^{-\nu} = 1/2$.
For a given network, the change in the synchronization error $\Omega_{\bar\varepsilon}(L)$ with respect to $\bar\varepsilon$ can be understood as the change in the value of the error function $\tilde\Omega(X)$ as $X = \bar\varepsilon\tilde{L}$ moves along a straight line determined by $\tilde{L}$.
From Eq.~\beqref{eqn:omega_eps2}, we expect in general that $\tilde\Omega(X)$ is a monotonically increasing function of $\rho(\tilde{L}^* - X)$, and hence $\Omega_{\bar\varepsilon}(L)$ is expected to decrease to a minimum at some $\bar\varepsilon = \bar\varepsilon^*$ and increases monotonically for $\bar\varepsilon > \bar\varepsilon^*$, or monotonically decrease in the entire range of $\bar\varepsilon$ for which $\Omega_{\bar\varepsilon}(L)$ is finite.
This is indeed the case for the example considered here, as illustrated by the insets in Fig.~\ref{figS1}.
Thus, in general we define $\Omega(L) := \inf_{\bar\varepsilon} \Omega_{\bar\varepsilon}(L)$ as a measure of synchronizability of a network, as it gives the lower limit on the asymptotic synchronization error for non-identical units.
For undirected networks, for which $\tilde{L}$ is symmetric and each $\lambda_i$ is real, diagonalization of $\tilde{L}$ with orthogonal eigenvectors can be used to show that
\begin{multline}
\Omega_{\bar\varepsilon}(L) = \sigma_\mu \tilde\Omega(\bar\varepsilon\tilde{L})
= \sigma_\mu \max_{2\le i \le n} \tilde\Omega^s (\bar\varepsilon\lambda_i), \quad\text{where } \\
\tilde\Omega^s(a) = \limsup_{T\to\infty} \Biggl\lvert\sum_{t=1}^T \left(\prod_{k=1}^{t-1}a_{k+T-t}\right)
b_{T-t} (1-a)^t \Biggr\rvert
\end{multline}
under the stability condition $\Lambda(\bar\varepsilon\lambda_i) < 0$ for $i = 2,\ldots,n$.
For such networks, the synchronization error can be determined visually from the error function $\tilde\Omega^s(a)$, which is a function of real numbers.
This is illustrated in Fig.~\ref{figS2} using the example of $F(x,\mu) = 2x + \mu \mod 1$, for which we can show that $\tilde\Omega^s(a) = \bigl\lvert\frac{1-a}{2a-1}\bigr\rvert$ and $\Lambda(a) = \ln 2 + \ln|1-a|$.

We now show that the class of networks with zero synchronization error for arbitrary heterogeneity of the individual units consists of those that are optimal [i.e., satisfies $\lambda_2 = \cdots = \lambda_n = \bar\lambda > 0$, Eq.~\textbf{3} in the main text] and have diagonalizable Laplacian matrix.
First, to show that any network with zero synchronization error satisfies these conditions, suppose
that $\Omega(L) = 0$ for a given network. 
That is, for some $\bar\varepsilon$, we have $\delta\mathbf{x}(t) \to 0$ as $t\to\infty$ for arbitrary $\mu_1,\ldots,\mu_n$ with a given $\sigma_\mu$.
Then, letting $t \to\infty$ in Eq.~\beqref{eqn:variational2}, we conclude that we either have ${\displaystyle \lim_{t\to\infty}} b_t = {\displaystyle\lim_{t\to\infty}} \frac{\del F}{\del \mu}(\bar{x}(t), \bar\mu) = 0$ or $(\tilde{L}^* -\bar\varepsilon \tilde{L})\delta\boldsymbol\mu = \mathbf{0}$ for all possible $\delta\boldsymbol\mu$.
The former can hold only in exceptional cases, such as when $\bar{x}(t)$ converges to a fixed point at which $\frac{\del F}{\del \mu}$ is zero.
We thus assume a typical situation in which the latter holds.
In this case, using the fact that the row sum of $\tilde{L}$ is zero and that $\sum_i\delta\mu_i = 0$, we can show that $\tilde{L}^* -\bar\varepsilon \tilde{L}$ must be equal to the zero matrix, and hence $\tilde{L} = \frac{1}{\bar\varepsilon}\tilde{L}^*$, which is diagonalizable with eigenvalues $0, \frac{1}{\bar\varepsilon}, \ldots, \frac{1}{\bar\varepsilon}$.
Since in general $L$ and $\tilde{L}$ have the same set of eigenvalues and $L$ is diagonalizable iff $\tilde{L}$ is diagonalizable, $L$ is diagonalizable and satisfies $\lambda_2 = \cdots = \lambda_n = \bar\lambda > 0$.

Conversely, suppose that the network satisfies $\lambda_2 = \cdots = \lambda_n = \bar\lambda > 0$ and the Laplacian matrix is diagonalizable.
It can be shown that $\bar\varepsilon\tilde{L} = \tilde{L}^*$ if we choose $\bar\varepsilon = 1/\bar\lambda$, and therefore $\delta\mathbf{x}(t) = \mathbf{0}$ according to Eq.~\beqref{eqn:variational2}, but we can actually prove stronger statement without the linear approximation involved in Eq.~\beqref{eqn:variational2}.
From Theorem~6 in Ref.~\cite{Nishikawa:2006kx2}, each node $j$ either has equal output link strength to all other nodes ($A_{ij} = b_j \neq 0$ for all $i \neq j$) or has no output at all ($A_{ij} = b_j = 0$ for all $i$).
This implies that the adjacency matrix satisfies $A_{ij} = b_j$ for all $i$ and $j$ with $i \neq j$, and we have $\sum_j b_j = \bar\lambda$.
If we choose $\bar\varepsilon = 1/\bar\lambda$, then Eq.~\beqref{HeteroEqn} becomes
\begin{equation}
x_i(t+1) = \frac{1}{\bar\lambda}\sum_{j=1}^n b_j F(x_j(t), \mu_j) = \sum_{j=1}^n w_j F(x_j(t), \mu_j),
\end{equation}
with $\sum_j w_j = 1$. 
Thus, the state of node $i$ is determined by the weighted average of the signals from all the nodes that have output and, more importantly, it is independent of $i$ for all $t \ge 1$, implying that $\delta\mathbf{x}(t) = \mathbf{0}$ for all $t \ge 1$.
Therefore, the system synchronizes in one iteration with zero error, despite the presence of dynamical heterogeneity, and hence $\Omega(L) = 0$.
This indicates that the largest Lyapunov exponent for the completely synchronous state is $-\infty$, which is analogous to super-stable fixed points and periodic orbits observed in maps.

Since the best networks for synchronizing non-identical maps satisfy $\lambda_2 = \cdots = \lambda_n = \bar\lambda > 0$, they too must have a quantized number of links: $m = k(n-1)$.
For every $n$ and every $k = 1,\ldots,n$, there is exactly one binary network ($A_{ij} = 0,1$) that has $m = k(n-1)$ links and is capable of complete synchronization for non-identical maps, including the directed star topology ($k=1$) and the fully connected network ($k=n$).
Note also that the above argument does not require that $b_j\ge 0$ for all $j$ (Theorem~6 in Ref.~\cite{Nishikawa:2006kx2} remains valid without this requirement).
This implies that complete synchronization is possible even for networks with negative interactions. 
In addition, the ability of a network to completely synchronize non-identical units, with or without negative interactions, is invariant under the generalized complement transformation defined by Eq.~\textbf{7} in the main text.
To see this, suppose that for a given network we have $\lambda_2 = \cdots = \lambda_n = \bar\lambda > 0$ and $L$ is diagonalizable.
By Theorem~6 in Ref.~\cite{Nishikawa:2006kx2}, we have $A_{ij} = b_j$ for all $i$ and $j$ with $i \neq j$, where $\sum_j b_j = \bar\lambda$.
Using the definition of the complement transformation, we have $A^c_{ij} = \alpha - b_j$, and $\sum_j (\alpha - b_j) = n\alpha - \bar\lambda > 0$ if $\alpha > \frac{m}{n(n-1)}$.
Applying Theorem~6 in Ref.~\cite{Nishikawa:2006kx2} again, we see that the complement satisfies the same property: $\lambda_2 = \cdots = \lambda_n = n\alpha - \bar\lambda > 0$ and its Laplacian matrix is diagonalizable.
Therefore, in addition to binary networks, there are many networks with negative interactions that are guaranteed to have zero synchronization error.

\section{5. Degree distribution before and after enhancing \\[-2mm]
synchronization with negative directional interactions}

\begin{figure}[t]
\begin{center}
\includegraphics[width=3.4in]{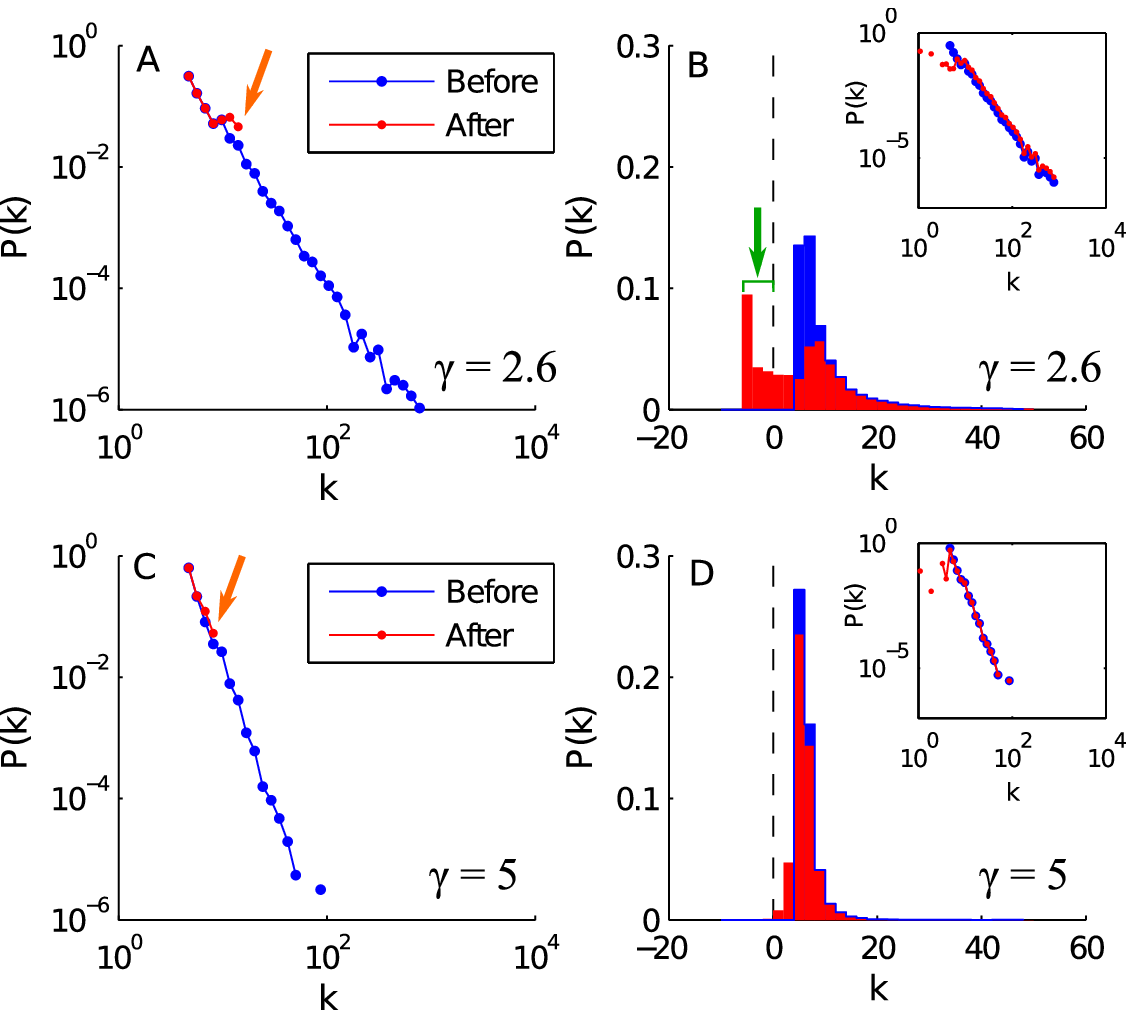}\rule{.3in}{0in}
\end{center}
\caption{\label{figS3}
Change in the degree distributions when enhancing synchronization with negative directional interactions.
Distributions before (blue) and after (red) assigning negative strengths 
in random scale-free networks:
(\textit{A}) in-degree distribution for $\gamma = 2.6$,
(\textit{B}) out-degree distribution for $\gamma = 2.6$,
(\textit{C}) in-degree distribution for $\gamma = 5$, and
(\textit{D}) out-degree distribution for $\gamma = 5$.
The in-degree distributions are plotted using logarithmic binning, and the absence of a symbol implies that we observed no node with its degree in the corresponding bin.
The out-degree distributions are shown in linear scale, along with insets showing the positive part of the distributions in logarithmic scale.
Note that there is a significant number of nodes with negative out-degree (green arrow).
All plots are averaged over 20 network realizations with 1,000 nodes.}
\end{figure}

We describe the change in the in- and out-degree distributions of the network as 
negative strengths are assigned to directional links to enhance synchronization, following the algorithm presented in the main text.
The in- and out-degree of node $i$ are defined as $\sum_{k\neq i} A_{ik}$ and $\sum_{k\neq i} A_{ki}$, respectively.
Figure~\ref{figS3} shows the results for random scale-free networks with $\gamma = 2.6$ and $\gamma = 5$.
They clearly illustrate that the large in-degree of many nodes are compensated by the negative interactions, creating a sharp cut-off in the distribution (orange arrows in panels \textit{A} and \textit{C}).
In contrast, the out-degree distributions remain essentially unchanged, having a power-law tail with the same exponent (insets in panels \textit{B} and \textit{D}).
Note that the algorithm can create negative out-degree nodes, as indicated by the green arrow in panel \textit{B}, but this has no significant effect since the in-degree distribution is the main factor that determines the stability of synchronous states.
\begin{figure}[t]
\begin{center}
\includegraphics[width=3.4in]{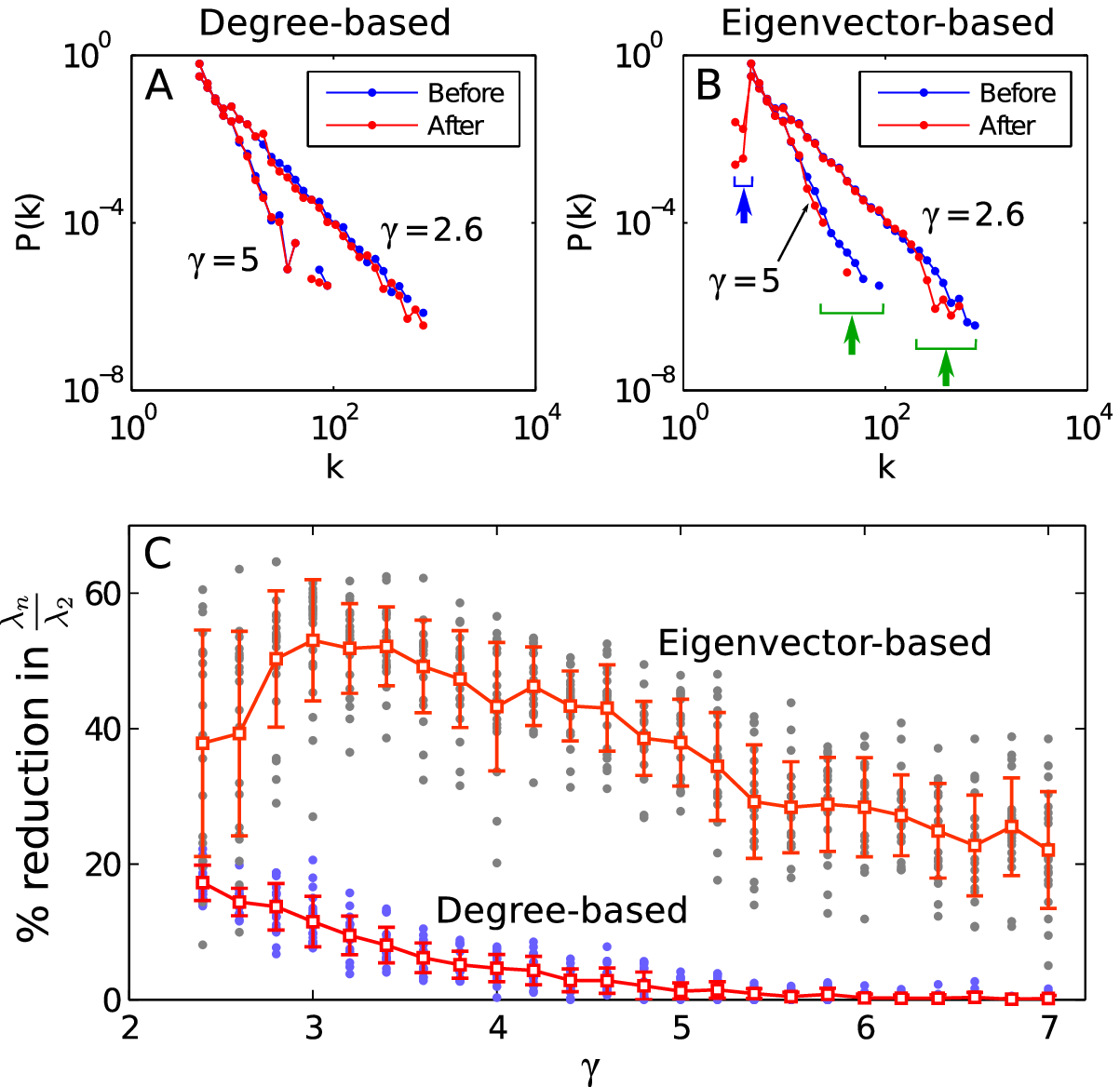}
\end{center}
\caption{\label{figS4}
Improving synchronization with negative bidirectional interactions. 
(\textit{A} and \textit{B}) 
Degree distributions before and after we apply two methods of assigning negative interactions to random scale-free networks with $\gamma = 2.6$ and 5:
(\textit{A}) degree-based method and
(\textit{B}) eigenvector-based method.
All four plots are generated using logarithmic binning and averaged over 20 network realizations with 1,000 nodes.
The absence of a symbol indicates that we observed no node with its degree in the corresponding bin.
(\textit{C}) Reduction in $\lambda_n/\lambda_2$ as a function of $\gamma$ for the two methods.
The error bars indicate the average and the standard deviation over 20 network realizations with 1,000 nodes. The individual realizations are indicated by gray and blue dots.}
\end{figure}

\newpage
\section{6. Enhancing synchronization with negative 
bidirec-\\[-2mm]tional interactions}

Here we show that 
assigning negative strength to bidirectional links
can also enhance synchronization significantly.
This is implemented using two different algorithms.

The first method is fast and is based on node degrees, similarly to the algorithm used in the main text for assigning negative directional interactions.
In order to create
negative interactions preferentially between nodes of large degrees, we first
order the bidirectional links according to the product of the degrees of the two nodes connected by each link, from high to low values.
Going through all the links in this order,
we change the strength of each bidirectional link from $+1$ to $-1$ if the degrees of the two adjacent nodes do not fall below a constant, chosen here arbitrarily to be 1.7 times the mean degree of the initial network.
We applied this procedure to random scale-free networks with minimum degree 5, generated by the configuration model.
Figure~\ref{figS4}\textit{A} shows the degree distribution before and after 
assigning
negative interactions for the scaling exponent $\gamma = 2.6$ and $5$.
In both cases, the degree distribution remains essentially scale-free with the same exponents.
Denoting by $\lambda_2$ and $\lambda_n$ the smallest and largest non-identically zero eigenvalues of the Laplacian matrix $L$, respectively, we measure synchronization enhancement by the relative decrease in the ratio $\lambda_n/\lambda_2$, a standard measure of synchronization widely adopted for undirected networks~\cite{Barahona:2002uo}.
Figure~\ref{figS4}\textit{C} shows that our method does produce significant enhancement for $\gamma$ less than about 5 (bottom curve), and that the more heterogeneous the initial degree distribution (smaller $\gamma$), the more effective the algorithm.

The second algorithm is slower but more effective, and it is based on the observation that the largest eigenvalue $\lambda_n$ is typically the one responsible for poor synchronizability in the undirected networks with heterogeneous degree distribution considered here~\cite{kim:248701}.
At each step of this algorithm, we use the first-order approximation~\cite{Hagberg:2008wd2,PhysRevE.81.046112} for each bidirectional link to estimate the change in $\lambda_n$ that would be caused by changing the strength of that link from $+1$ to $-1$.
We then choose a link with the largest predicted reduction in $\lambda_n$ and make its strength $-1$.
Repeating this until the fraction of links with negative strength reaches a prescribed threshold (chosen here arbitrarily to be 0.2), we obtain a sequence of candidate networks for improved synchronization\footnote{Although we have chosen to reduce $\lambda_n$ here, a similar algorithm can be conceived to increase $\lambda_2$ or to simultaneously optimize $\lambda_2$ and $\lambda_n$.}.
From these networks we choose one with the smallest ratio $\lambda_n/\lambda_2$ as the output of the algorithm.
Figure~\ref{figS4}\textit{B} shows the change in the degree distribution before and after applying this algorithm to random scale-free networks with minimum degree 5 for $\gamma = 2.6$ and $5$, generated by the configuration model.
For both values of $\gamma$, we notice a significant drop in the fraction of high-degree nodes (green arrows), accompanied by the appearance of nodes with degree less than the minimum degree of the initial network (blue arrow).
This change appears to be responsible for the reduction of $\lambda_n/\lambda_2$ by as much as about 65\%, as shown in Fig.~\ref{figS4}\textit{C} (top curve).
Note that the synchronization enhancement achieved by this method is consistently larger than the first method based on link degrees (bottom curve).
The effectiveness of this method depends on the fact that the Laplacian eigenvalues remain real when the network is kept symmetric, which would not generally hold true if 
negative strength were assigned to directional links.

\section*{Supporting Video}
\verb#http://www.youtube.com/watch?v=3dMI1Yyxmbw#\\[2mm]
Networks with best synchronization. The first half of the movie shows the structural changes in networks with best synchronization properties (smallest $\sigma$ possible for a given number of links) as directional links are removed one by one.
We always choose a link that keeps the synchronizability highest (i.e., keeps $\sigma$ smallest).  The second half shows how $\sigma$ changes in the process, revealing in particular that link removal can counterintuitively enhance synchronization.  The node layout at each step was computed using the Kodama-Kawai spring layout algorithm~\cite{kodama-kawai}.
We used the implementation of the algorithm in the Boost C++ library~\cite{boost} through the Matlab interface provided by MatlabBGL~\cite{Matlab_bgl}.

\begin{center}
\includegraphics[width=2in]{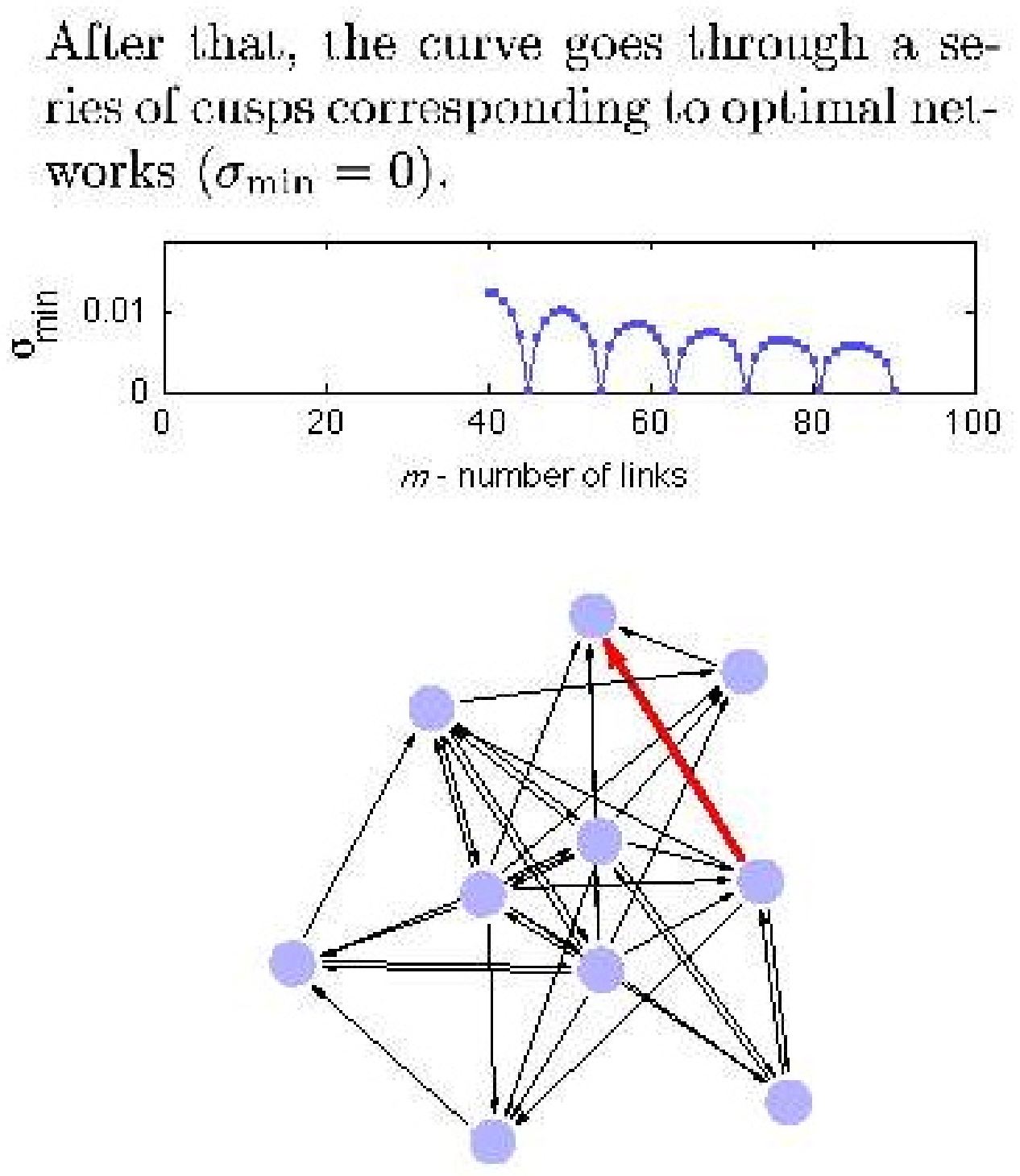}
\end{center}


\end{article}

\end{document}